\documentclass[10pt,conference]{IEEEtran}
\IEEEoverridecommandlockouts
\usepackage{cite}
\usepackage{amsmath,amssymb,amsfonts}
\usepackage{bm}
\usepackage{algorithmic}
\usepackage[pdftex]{graphicx}
\usepackage{textcomp}
\usepackage{multirow}
\usepackage{xcolor}
\def\BibTeX{{\rm B\kern-.05em{\sc i\kern-.025em b}\kern-.08em
    T\kern-.1667em\lower.7ex\hbox{E}\kern-.125emX}}

\newcommand{\rbm}[1]{\bm{\mathrm{#1}}}

\newcommand{\diag}{\mathop{\mathrm{diag}}}
\newcommand{\HH}{\mathop{\textsf{H}}}
\renewcommand{\top}{\mathop{\textsf{T}}}

\newcommand{\bhline}[1]{\noalign{\hrule height #1}}
\newcommand{\tr}{\mathop{\mathrm{tr}}}

\newcommand{\xij}{\bm{x}_{ij}}

\newcommand{\trh}{\tilde{r}_{ij}^{(h)}}

\newcommand{\ai}{\bm{a}_{i}^{(h)}}

\newcommand{\rijx}{\rbm{R}_{ij}^{(x)}}

\begin{document}
\setlength{\abovedisplayskip}{6pt}
\setlength{\belowdisplayskip}{6pt}
\allowdisplaybreaks[1]

\title{Efficient Full-Rank Spatial Covariance Estimation Using Independent Low-Rank Matrix Analysis\\ for Blind Source Separation}

\author{\IEEEauthorblockN{Yuki Kubo\IEEEauthorrefmark{2}, Norihiro Takamune\IEEEauthorrefmark{2},
Daichi Kitamura\IEEEauthorrefmark{3}, Hiroshi Saruwatari\IEEEauthorrefmark{2}}
\IEEEauthorblockA{\IEEEauthorrefmark{2}\textit{The University of Tokyo, Graduate School of Information Science and Technology,}\\
7-3-1 Hongo, Bunkyo-ku, Tokyo 113-8656, Japan}
\IEEEauthorblockA{\IEEEauthorrefmark{3}\textit{National Institute of Technology, Kagawa College,}\\
355 Chokushi-cho, Takamatsu, Kagawa 761-8058, Japan}
}

\maketitle

\begin{abstract}
In this paper, we propose a new algorithm that efficiently separates a directional source and diffuse background noise based on independent low-rank matrix analysis (ILRMA).
ILRMA is one of the state-of-the-art techniques of blind source separation (BSS) and is based on a rank-1 spatial model.
Although such a model does not hold for diffuse noise, ILRMA can accurately estimate the spatial parameters of the directional source.
Motivated by this fact, we utilize these estimates to restore the lost spatial basis of diffuse noise, which can be considered as an efficient full-rank spatial covariance estimation.
BSS experiments show the efficacy of the proposed method in terms of the computational cost and separation performance.
\end{abstract}


\section{Introduction}
\label{sec:intro}
Blind source separation (BSS) is a technique for separating an observed multichannel signal, which is a mixture of multiple sources, into each source without any prior information about the sources or the mixing system.
In a determined or overdetermined situation (number of sensors $\geq$ number of sources), frequency-domain independent component analysis (FDICA) \cite{PSmaragdis1998_BSS,HSaruwatari2006_FDICA}, independent vector analysis (IVA) \cite{AHiroe2006_IVA,TKim2007_IVA}, and independent low-rank matrix analysis (ILRMA) \cite{DKitamura2016_ILRMA,DKitamura2018_ILRMA} have been proposed for audio BSS problems.
In particular, ILRMA assumes low-rankness for the power spectrogram of each source using nonnegative matrix factorization (NMF) \cite{DDLee1999_NMF,DDLee2000_NMF} in addition to statistical independence between sources, and achieves efficient and accurate separation \cite{DKitamura2016_ILRMA}.
These methods assume a rank-1 spatial model; the frequency-wise acoustic path of each source can be represented by a single time-invariant spatial basis, which is often called a steering vector.
Under this assumption, the determined BSS problem reduces to the estimation of a demixing matrix for each frequency.
However, the assumption in the rank-1 spatial model becomes invalid in actual situations.
For instance, when a target source (directional source) and diffuse noise that arrives from all directions are mixed, FDICA, IVA, and ILRMA cannot extract only the target source in principle \cite{SAraki2003_ICAeqBF}, and the estimated target source includes residual diffuse noise.

Multichannel NMF (MNMF) \cite{AOzerov2010_MNMF,HSawada2013_MNMF} is theoretically equivalent to ILRMA except for the mixing model, namely, MNMF employs a full-rank spatial covariance matrix \cite{Duong2010_SpatialMatrix}.
This model can represent not only the acoustic path but also the spatial spread of each source or diffuse noise, while its optimization has a huge computational cost and lacks robustness against the initialization \cite{DKitamura2016_ILRMA}.
To accelerate the parameter estimation, FastMNMF has been proposed~\cite{Ito2019_FastMNMF, Sekiguchi2019_FastMNMF}.
It assumes a jointly diagonalizable spatial covariance matrix to greatly reduce the computational cost of the update algorithm,
although its performance still depends on the initial values of parameters.
To increase the stability of its performance, ILRMA-based initialization was utilized for MNMF in \cite{Shimada2018UnsupervisedBB}.
However, the improvement is still limited because of the complexity of optimization with a large number of parameters.

In this paper, we treat the BSS problem with one directional target source and diffuse background noise, where more than or equal to two microphones are available.
In this case, the target source can be expressed using the rank-1 spatial covariance (one steering vector), but diffuse noise requires the full-rank spatial covariance because of its spatial spread.
To achieve robust and computationally efficient BSS in this situation, we propose a new approach based on ILRMA: (a) rank-1 target covariance and rank-$(M\!-\!1)$ diffuse noise covariance matrices are simultaneously estimated by ILRMA, where $M$ is the number of microphones,
(b) one lost spatial basis for diffuse noise is restored to obtain the rank-$M$ (full-rank) noise covariance via the expectation-maximization (EM) algorithm, and (c) a multichannel Wiener filter is applied to enhance only the target source.
The efficacy of the proposed method is confirmed through BSS experiments using a mixture of speech and diffuse noise.

Regarding its relation to prior works, the proposed method is considered as a spatial model extension of FDICA, IVA, and ILRMA, which are the conventional independence-based BSS algorithms utilizing the rank-1 spatial model.
Compared with conventional MNMF and FastMNMF based on the full-rank spatial model, the proposed method is regarded as a computationally efficient algorithm with higher separation accuracy.

\section{Independent Low-Rank Matrix Analysis}
\subsection{Formulation}
\label{ssec:ILRMAformulation}
Let us denote a multichannel observed signal as $\bm{x}_{ij}=(x_{ij,1}, \dots, x_{ij,m}, \ldots, x_{ij,M})^{\top}\in\mathbb{C}^{M}$ that is obtained via a short-time Fourier transform (STFT), where $i=1, \dots, I$, $j=1, \dots, J$, and $m=1, \dots, M$ are the indices of the frequency bins, time frames, and microphones, respectively, and $^{\top}$ denotes the transpose.
Also, source signals (dry sources) are denoted as $\bm{s}_{ij}=(s_{ij,1}, \dots, s_{ij,n}, \dots, s_{ij,N})^{\top}\in\mathbb{C}^{N}$, where $n=1, \dots, N$ is the index of the sources and $N$ is the number of sources.
If each source in $\bm{x}_{ij}$ can be represented by a time-invariant steering vector $\bm{a}_{i,n}\in\mathbb{C}^{M}$, the following mixing system holds:
\begin{equation}
\label{eq:mixgen}
	\bm{x}_{ij}=\rbm{A}_{i}\bm{s}_{ij},
\end{equation}
where $\rbm{A}_{i}=(\bm{a}_{i,1}\ \cdots\ \bm{a}_{i,N})$ is called a mixing matrix.
If $M=N$ and $\rbm{A}_{i}$ is invertible, the separated signal $\bm{y}_{ij} = (y_{ij,1}, \dots, y_{ij,N})^{\top}\in\mathbb{C}^{N}$ can be obtained by estimating the demixing matrix $\rbm{W}_{i}=(\bm{w}_{i,1}\ \cdots\ \bm{w}_{i,N})^{\HH} = \mathbf{A}_{i}^{-1}$ as 
\begin{equation}
\bm{y}_{ij}=\rbm{W}_{i}\bm{x}_{ij},
\end{equation}
where $ ^{\HH}$ denotes the Hermitian transpose.

\subsection{Generative Model and Update Rules}
\label{ssec:ILRMAestimation}

In ILRMA, as the generative model of source signals, the following complex Gaussian distribution is assumed:
\begin{equation}
\label{eq:ILRMAsrc}
s_{ij,n}\sim\mathcal{N}_c\left(0,r_{ij,n}\right),
\end{equation}
where $r_{ij,n}$ is the time-frequency-varying variance (power spectrogram model of $s_{ij,n}$).
Also, $r_{ij,n}$ is modeled by NMF \cite{CFevotte2009_ISNMF} as $r_{ij,n}=\sum_{l}t_{il,n}v_{lj,n}$, where $t_{il,n}\geq 0$ and $v_{lj,n}\geq 0$ are the NMF variables, $l=1, \dots, L$ is the index of the NMF bases, and $L$ is the number of bases.
From (\ref{eq:mixgen}) and (\ref{eq:ILRMAsrc}), the generative model of the observed signal becomes
\begin{align}
\label{eq:ILRMAobs}
\bm{x}_{ij}&\sim\mathcal{N}_c\left(\bm{0},\sum_{n}r_{ij,n}\bm{a}_{i,n}\bm{a}_{i,n}^{\HH}\right).
\end{align}
Since the mixing system (\ref{eq:mixgen}) is assumed in ILRMA, the spatial covariance is represented by a rank-1 matrix as $\bm{a}_{i,n}\bm{a}_{i,n}^{\HH}$, which is called the rank-1 spatial model.

The cost function in ILRMA is defined as the negative log-likelihood function of (\ref{eq:ILRMAobs}) as 
\begin{equation}
\label{eq:ILRMAcost}
\mathcal{L} = -2J\sum_{i}\log{|\det\rbm{W}_{i}|}+\sum_{i,j,n}\left(\frac{|y_{ij,n}|^2}{r_{ij,n}}+\log r_{ij,n}\right),
\end{equation}
where $y_{ij,n}=\bm{w}_{i,n}^{\HH}\bm{x}_{ij}$.
Both the separation filter $\bm{w}_{i,n}$ and the NMF variables $t_{il,n}$ and $v_{lj,n}$ can be optimized in the maximum likelihood sense (minimization of (\ref{eq:ILRMAcost})) by iterating the following iterative update rules \cite{DKitamura2016_ILRMA}:
\begin{align}
\rbm{G}_{i,n}&=\frac{1}{J}\sum_{j}\frac{1}{r_{ij,n}}\bm{x}_{ij}\bm{x}_{ij}^{\HH},\\
\bm{w}_{i,n}&\leftarrow(\rbm{W}_{i}\rbm{G}_{i,n})^{-1}\bm{e}_{n},\\
\bm{w}_{i,n}&\leftarrow\bm{w}_{i,n}(\bm{w}_{i,n}^{\HH}\rbm{G}_{i,n}\bm{w}_{i,n})^{-\frac{1}{2}},
\end{align}
where $\bm{e}_{n}$ denotes the unit vector with the $n$th element equal to unity.
The update rules for $\bm{w}_{i,n}$ are called the iterative projection~\cite{NOno2011_AuxIVA}, which promises convergence-guaranteed efficient optimization.
Also, we can update $t_{il,n}$ and $v_{lj,n}$ by minimizing the Itakura--Saito divergence between $\sum_{l}t_{il,n}v_{lj,n}$ and $r_{ij,n}$ (see \cite{DKitamura2016_ILRMA} for details).

\section{Proposed Method}
\subsection{Motivation and Strategy}
\label{ssec:motiv}
In this paper, we deal with a mixture signal that includes one directional target source and diffuse background noise.
Since diffuse noise cannot be expressed by the rank-1 spatial model (one steering vector),
BSS based on a full-rank covariance model, such as MNMF, should be applied in this situation.
However, estimation of the full-rank covariance has a huge computational cost,
and its performance is always more unstable than ILRMA~\cite{DKitamura2016_ILRMA}
because of the large number of spatial parameters, $INM^2$, which can be reduced to $INM$ using the rank-1 spatial model (ILRMA).

\begin{figure}[t]
\centering\includegraphics[width=\linewidth]{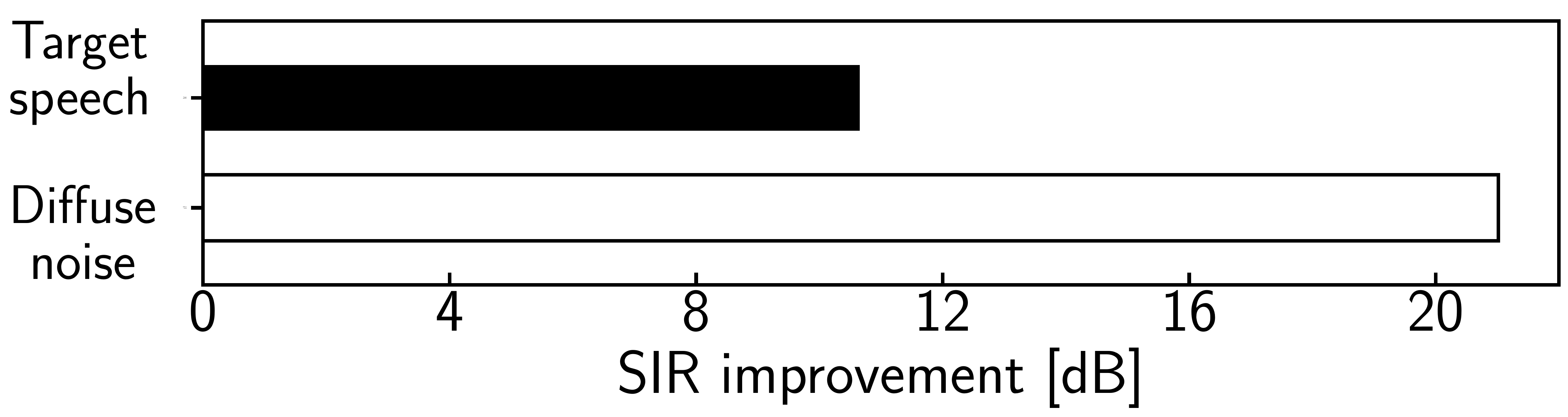}
\vspace{-5mm}
\caption{SIR improvement for directional speech and diffuse noise.}
\label{fig:SIRilrma}
\vspace{-3mm}
\end{figure}

For this reason, to achieve efficient and stable BSS, we propose a new ILRMA-based full-rank covariance estimation using more than or equal to two microphones.
Although the sources are categorized into two groups (target and noise), we assume that one target source and $M-1$ noise components are mixed ($N=M$).
This assumption allows us to model the diffuse noise using $M-1$ spatial bases (rank-$(M\!-\!1)$ spatial covariance).
The extraction of the target source in this manner is still 
difficult because noise components exist even in the same direction as the target source.
However, FDICA or ILRMA can separate the diffuse noise with high accuracy even if one spatial basis for diffuse noise is lacking.
Figure \ref{fig:SIRilrma} shows an example of the separation performance (source-to-interference ratio (SIR) \cite{EVincent2006_BSSEval}) obtained by ILRMA,
where directional speech and diffuse noise are mixed and the experimental conditions are described in Sect.~\ref{sec:exp}.
It can be seen that diffuse noise is accurately estimated (almost perfectly with more than 20 dB accuracy) rather than the target speech, where diffuse noise is modeled using the rank-($M\!-\!1$) spatial covariance.
This is because the demixing filters for the diffuse noise can precisely cancel the target speech, which is a point source \cite{YTakahashi2009_BSSA}, meaning that the steering vector of the directional source $\bm{a}_{i,n_h}$ can be estimated by ILRMA with high accuracy, where $n_h$ denotes the index of the target source.
This implies that we can fix some spatial parameters in the full-rank spatial model for diffuse noise by utilizing the estimates obtained by ILRMA in advance.

On the basis of the above motivation, we propose the following new estimation method for the full-rank spatial covariance of diffuse noise:
(a) the rank-1 spatial covariance for the target source, $\bm{a}_{i,n_h}\bm{a}_{i,n_h}^{\HH}$, and rank-$(M\!-\!1)$ covariance for diffuse noise,
$\sum_{n\neq n_h}\bm{a}_{i,n}\bm{a}_{i,n}^{\HH}$, are estimated by ILRMA, 
(b) the lost spatial basis for diffuse noise is restored via the EM algorithm to estimate the noise components in the direction of the target source,
and (c) a multichannel Wiener filter is applied to suppress the noise components remaining in the separated target source.

\subsection{Model of Target Source and Diffuse Noise}
The observed signal $\bm{x}_{ij}$ is assumed to be the sum of two components, as 
\begin{equation}
\bm{x}_{ij}=\bm{h}_{ij}+\bm{u}_{ij},
\end{equation}
where $\bm{h}_{ij}=(h_{ij,1},\ldots,h_{ij,M})^{\top}\in\mathbb{C}^M$ is the spatial image of the target source and $\bm{u}_{ij}=(u_{ij,1},\ldots,u_{ij,M})^{\top}\in\mathbb{C}^M$ is that of the diffuse noise.
The target source $\bm{h}_{ij}$ is modeled as 
\begin{align}
\bm{h}_{ij}&=\bm{a}_{i}^{(h)}s_{ij}^{(h)},\\
s_{ij}^{(h)}&\sim\mathcal{N}_c(0,r_{ij}^{(h)}),\label{eq:targetsrc}
\end{align}
where $\bm{a}_{i}^{(h)}$, $s_{ij}^{(h)}$, and $r_{ij}^{(h)}$ are the $n_h$th steering vector $\bm{a}_{i,n_h}$, the dry source component,
and the power spectrogram of the $n_h$th source, respectively.
As mentioned in Sect.~\ref{ssec:motiv}, $\bm{a}_{i}^{(h)}$ can be accurately estimated by ILRMA.
Thus, we hereafter consider $\bm{a}_{i}^{(h)}$ as a given and fixed parameter in the following processes.
In addition to (\ref{eq:targetsrc}), to improve the estimation performance, we introduce an a priori distribution for the variance $r_{ij}^{(h)}$ using the inverse gamma distribution,
\begin{equation}
p(r_{ij}^{(h)};\alpha,\beta)=\frac{\beta^{\alpha}}{\Gamma(\alpha)}\left(r_{ij}^{(h)}\right)^{-\alpha-1}\exp\left(-\frac{\beta}{r_{ij}^{(h)}}\right),
\end{equation}
where $\alpha>0$ and $\beta>0$ are shape and scale parameters, respectively,
and a large $\alpha$ with a small $\beta$ induces the sparseness of $r_{ij}^{(h)}$.

Since diffuse noise should have a full-rank spatial covariance, the generative model of $\bm{u}_{ij}$ is expressed by a multivariate complex Gaussian distribution as 
\vspace{0mm}
\begin{equation}
\bm{u}_{ij}\sim\mathcal{N}_c(\bm{0},r_{ij}^{(u)}\rbm{R}_{i}^{(u)}),
\end{equation}
where $r_{ij}^{(u)}$ and $\rbm{R}_{i}^{(u)}$ are the variance and spatial covariance for the diffuse noise, respectively.
From the estimated demixing filter $\bm{w}_{i,n}$ obtained by ILRMA, we can model the full-rank spatial covariance of the diffuse noise as follows:
\begin{align}
\rbm{R}_{i}^{(u)}&=\rbm{R}_{i}'^{(u)}+\lambda_{i}\bm{b}_{i}\bm{b}_{i}^{\HH},\\
\mathbf{R}_{i}'^{(u)} &= \frac{1}{J}\sum_{j}\rbm{W}_{i}^{-1}\diag\left(|\bm{w}_{i,1}^{\HH}\bm{x}_{ij}|^2, \dots, |\bm{w}_{i,n_h-1}^{\HH}\bm{x}_{ij}|^2, 0,\right.\nonumber\\
&\phantom{=}\left.|\bm{w}_{i,n_h+1}^{\HH}\bm{x}_{ij}|^2, \dots, |\bm{w}_{i,N}^{\HH}\bm{x}_{ij}|^2\right)\left(\rbm{W}_{i}^{-1}\right)^{\HH},\label{eq:Rprime}
\end{align}
where $\bm{b}_{i}$ is the unit eigenvector of $\rbm{R}_{i}'^{(u)}$ that corresponds to the zero eigenvalue
and $\lambda_{i}$ is a scalar weight used to complement the lost spatial basis, namely, the direction of the target source.
Note that (\ref{eq:Rprime}) includes a back-projection operation to compensate the scales of the signals \cite{NMurata2001_Permutation}.
Since $\rbm{R}_{i}'^{(u)}$ consists of $M-1$ noise estimates, its rank is $M-1$.
Therefore, to restore the lost spatial basis in $\rbm{R}_{i}'^{(u)}$, we must simultaneously estimate the eigenvalue $\lambda_{i}$, the variance of the target source $r_{ij}^{(h)}$, and the variance of the diffuse noise $r_{ij}^{(u)}$ with $\bm{a}_{i}^{(h)}$ and the rank-$(M\!-\!1)$ spatial covariance $\rbm{R}_{i}'^{(u)}$ fixed.
In summary, the number of spatial parameters to be estimated in the proposed method is $INM$ (for ILRMA) $\mbox{}+I$ (for $\lambda_{i}$), i.e., $I(NM+1)$, which is much less than that of MNMF ($INM^2$) and FastMNMF ($IM^2+INM$).
\subsection{Update Rules Based on EM Algorithm}
The parameters $\lambda_{i}$, $r_{ij}^{(h)}$, and $r_{ij}^{(u)}$ are optimized by a maximum a posteriori estimation based on the EM algorithm.
A $Q$ function is defined by the expected value of the complete-data log-likelihood w.r.t. $p(s_{ij}^{(h)}, \bm{u}_{ij}|\bm{x}_{ij};\tilde{\Theta})$ as
\begin{align}
    Q(\Theta;\tilde{\Theta}) &= \sum_{i,j}\Biggl[-(\alpha+2)\log r_{ij}^{(h)}-\frac{\hat{r}_{ij}^{(h)}+\beta}{r_{ij}^{(h)}}-M\log r_{ij}^{(u)}\nonumber\\
    &\phantom{=}\biggl.\mbox{}-\log\det\rbm{R}_{i}^{(u)}-\frac{\tr\left((\rbm{R}_{i}^{(u)})^{-1}\hat{\rbm{R}}_{ij}^{(u)}\right)}{r_{ij}^{(u)}}\Biggr]+\mathrm{const.,}
\end{align}
where $\mathrm{const.}$ includes the constant terms that do not depend on the parameters, $\Theta=\left\{r_{ij}^{(h)},r_{ij}^{(u)},\lambda_i\right\}$ is the set of parameters to be updated, $\tilde{\Theta}=\left\{\tilde{r}_{ij}^{(h)},\tilde{r}_{ij}^{(u)},\tilde{\lambda}_{i}\right\}$ is the set of up-to-date parameters, and $\hat{r}_{ij}^{(h)}$ and $\hat{\rbm{R}}_{ij}^{(u)}$ are the sufficient statistics obtained by the E-step.
The update rules in the E-step are as follows:
\begin{align}
\tilde{\rbm{R}}_{i}^{(u)}&=\rbm{R}_{i}'^{(u)}+\tilde{\lambda}_{i}\bm{b}_{i}\bm{b}_{i}^{\HH},\\
\rbm{R}_{ij}^{(x)}&=\tilde{r}_{ij}^{(h)}\bm{a}_{i}^{(h)}\left(\bm{a}_{i}^{(h)}\right)^{\HH}+\tilde{r}_{ij}^{(u)}\tilde{\rbm{R}}_{i}^{(u)}\label{eq:defRx},\\
\hat{r}_{ij}^{(h)}&=\tilde{r}_{ij}^{(h)}-\left(\tilde{r}_{ij}^{(h)}\right)^2\left(\bm{a}_{i}^{(h)}\right)^{\HH}\left(\rbm{R}_{ij}^{(x)}\right)^{-1}\bm{a}_{i}^{(h)}\nonumber\\
    &\phantom{=}+\left|\trh\xij^{\HH}(\rijx)^{-1}\ai\right|^2,\\
\hat{\rbm{R}}_{ij}^{(u)}&=\tilde{r}_{ij}^{(u)}\tilde{\rbm{R}}_{i}^{(u)}-\left(\tilde{r}_{ij}^{(u)}\right)^2\tilde{\rbm{R}}_{i}^{(u)}\left(\rbm{R}_{ij}^{(x)}\right)^{-1}\tilde{\rbm{R}}_{i}^{(u)}\nonumber\\
&\phantom{=}+\left(\tilde{r}_{ij}^{(u)}\right)^2\tilde{\rbm{R}}_{i}^{(u)}\left(\rbm{R}_{ij}^{(x)}\right)^{-1}\bm{x}_{ij}\bm{x}_{ij}^{\HH}\left(\rbm{R}_{ij}^{(x)}\right)^{-1}\tilde{\rbm{R}}_{i}^{(u)}.
\end{align}
In the M-step, we employ a coordinate ascent algorithm to the $Q$ function.
The update rules are as follows:
\begin{align}
r_{ij}^{(h)}&\leftarrow \frac{\hat{r}_{ij}^{(h)}+\beta}{\alpha+2},\\
\rbm{K}_{i}&=\frac{1}{J}\sum_{j}\frac{1}{\tilde{r}_{ij}^{(u)}}\hat{\rbm{R}}_{ij}^{(u)},\\
\lambda_{i}&\leftarrow \bm{b}_{i}^{\HH}\rbm{K}_{i}\bm{b}_{i},\\
\rbm{R}_{i}^{(u)}&\leftarrow\rbm{R}_{i}'^{(u)}+\lambda_{i}\bm{b}_{i}\bm{b}_{i}^{\HH}\label{eq:estRu},\\
r_{ij}^{(u)}&\leftarrow\frac{1}{M}\tr\left(\left(\rbm{R}_{i}^{(u)}\right)^{-1}\hat{\rbm{R}}_{ij}^{(u)}\right).
\end{align}
\subsection{Multichannel Wiener Filter}
After the estimation of all the parameters, the following multichannel Wiener filter is employed:
\begin{align}
\hat{\bm{h}}_{ij}&=r_{ij}^{(h)}\bm{a}_{i}^{(h)}\left(\bm{a}_{i}^{(h)}\right)^{\HH}\left(\rbm{R}_{ij}^{(x)}\right)^{-1}\bm{x}_{ij},\\
\hat{\bm{u}}_{ij}&=r_{ij}^{(u)}\rbm{R}_{i}^{(u)}\left(\rbm{R}_{ij}^{(x)}\right)^{-1}\bm{x}_{ij}.
\end{align}

\subsection{Initialization of Source Variances}
Since the EM algorithm strongly depends on the initial values of the parameters, we employ the ILRMA estimates to initialize the source variances $r_{ij}^{(h)}$ and $r_{ij}^{(u)}$ to avoid trapping at a poor local solution as follows:
\begin{align}
r_{ij}^{(h)}&=\sum_{l}t_{il,n_h}v_{lj,n_h},\\
r_{ij}^{(u)}&=\frac{1}{M}\left(\hat{\bm{y}}_{ij}^{(u)}\right)^{\HH}\left(\rbm{R}_{i}'^{(u)}\right)^{+}\hat{\bm{y}}_{ij}^{(u)},
\end{align}
where $t_{il,n_h}$ and $v_{lj,n_h}$ are the low-rank source model of the target source obtained by ILRMA, $^{+}$ denotes the pseudoinverse, and $\hat{\bm{y}}_{ij}^{(u)}$ is the scale-fixed source image of diffuse noise obtained as $\sum_{n\neq n_h}\rbm{W}_{i}^{-1}(0, \dots, 0, \bm{w}_{i,n}^{\HH}\bm{x}_{ij}, 0, \dots, 0)^{\top}$.
Also, $\lambda_{i}$ is initialized by the minimum nonzero eigenvalue of $\rbm{R}_{i}'^{(u)}$.

\section{Experiments}
\label{sec:exp}
\begin{figure*}[t]
\setcounter{figure}{2}
\centering\includegraphics[width=0.95\linewidth]{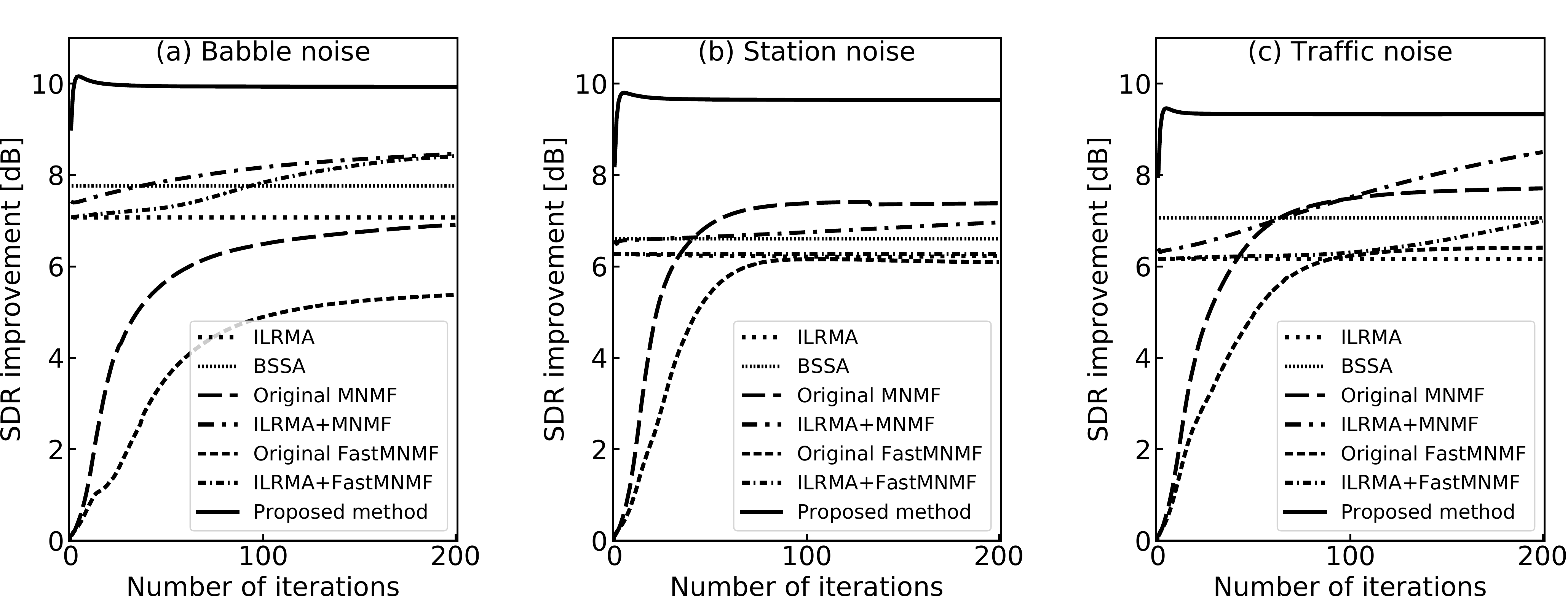}
\caption{\label{fig:SDRall} SDR behaviors averaged over 10 parameter-initialization random seeds and four target directions in separation of target speech and diffuse (a) babble, (b) station, (c) traffic noises, where speech-to-noise ratio is 0 dB.}
\end{figure*}
\begin{figure}[t]
\setcounter{figure}{1}
\centering
\includegraphics[width=5.47cm]{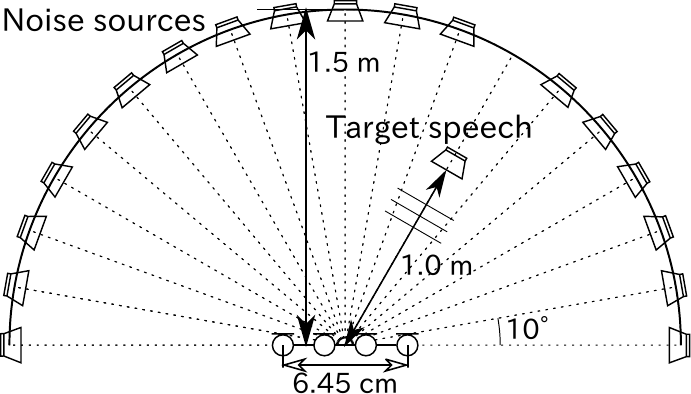}
    \caption{\label{fig:reccond}Recording conditions of impulse responses (when target source is located at $30^{\circ}$), where reverberation time $T_{60}$ is 200 ms.}
\end{figure}
\begin{table}
\centering
\caption{\label{tbl:expcond}Experimental conditions}
\small
\begin{tabular}{cc}\bhline{1.5pt}
Sampling frequency & 16 kHz\\\hline

\multirow{2}{*}{STFT} & 256-ms-long Hamming\\
& window with 128 ms shift\\\hline
Number of NMF bases $L$ & 10 for source model\\\hline
Number of iterations &\multirow{2}{*}{50}\\
in ILRMA & \\\hline
Number of iterations & \multirow{2}{*}{200}\\
    in methods except ILRMA\\\bhline{1.5pt}
\end{tabular}
\end{table}
\subsection{Experimental Conditions}
To confirm the efficacy of the proposed method, we conducted a BSS experiment using a simulated mixture of a target speech source and diffuse noise.
We compared seven methods, namely, ILRMA~\cite{DKitamura2016_ILRMA}, BSSA~\cite{YTakahashi2009_BSSA}, the original MNMF~\cite{HSawada2013_MNMF}, MNMF initialized by ILRMA (ILRMA+MNMF)~\cite{DKitamura2016_ILRMA,Shimada2018UnsupervisedBB}, the original FastMNMF~\cite{Sekiguchi2019_FastMNMF}, FastMNMF initialized by ILRMA (ILRMA+FastMNMF), and the proposed method ($\alpha=0.7$ and $\beta=10^{-16}$ were selected experimentally).
In ILRMA, the observation $\xij$ was preprocessed via a sphering transformation using PCA.
For BSSA, we replaced FDICA in \cite{YTakahashi2009_BSSA} with ILRMA and set the oversubtraction and flooring parameters to 1.4 and 0, respectively.
For ILRMA, the original MNMF, and the original FastMNMF, all the NMF variables were initialized by nonnegative random values.
The demixing matrix $\rbm{W}_{i}$ in ILRMA and the spatial covariance matrix in the original MNMF and the original FastMNMF were initialized by the identity matrix $\rbm{I}$.
For ILRMA+MNMF and ILRMA+FastMNMF, the NMF variables were taken from ILRMA.
Also, the spatial covariance matrix was initialized using $\bm{a}_{i,n}\bm{a}_{i,n}^{\HH}+\varepsilon\rbm{I}$ for ILRMA+MNMF and $\bm{a}_{i,n}\bm{a}_{i,n}^{\HH}+\varepsilon\sum_{n'\neq n}\bm{a}_{i,n'}\bm{a}_{i,n'}^{\HH}$ for ILRMA+FastMNMF, where $\bm{a}_{i,n}$ was estimated by ILRMA and $\varepsilon$ was set to $10^{-5}$.

We used speech signals obtained from the JNAS speech corpus \cite{KItou1999_JNAS} to produce the target speech source and diffuse babble noise.
The station and traffic noise signals were obtained from DEMAND \cite{Joachim2013_DEMAND}.
These dry sources were convoluted with the impulse responses shown in Fig.~\ref{fig:reccond} to simulate the mixture,
where the target source was located at $30^{\circ}$, $20^{\circ}$, $10^{\circ}$, or $0^{\circ}$ clockwise from the normal to a microphone array, the 18 loudspeakers used to simulate diffuse noise were arranged at intervals of $10^{\circ}$ except in the target source direction, the size of the recording room for these impulse responses was 3.9~m $\times$ 3.9~m, and its reverberation time was about 200 ms. 
Note that the diffuse babble noise was produced by convoluting 18 independent speakers with each impulse response, and the diffuse station and traffic noises were produced by splitting the dry source into 18 short-time periods and convoluting them with each impulse response.
The speech-to-noise ratio was set to 0 dB.
The other conditions are shown in Table~\ref{tbl:expcond}.

\subsection{Results}
\vspace{-1mm}
Source-to-distortion ratio (SDR)~\cite{EVincent2006_BSSEval} is used as a total evaluation score in terms of separation performance and sound distortion.
The SDR behaviors for each of the methods, which are the averaged results over 10 parameter-initialization random seeds and four target directions, are shown in Fig.~\ref{fig:SDRall},
where those of ILRMA-initialized methods are depicted except for their initializing iterations of ILRMA.
The proposed method outperformed the other methods.
In particular, the full-rank spatial model in the proposed method showed an improvement of more than 3~dB compared with the rank-1 spatial model in ILRMA, and the efficacy of the proposed spatial model extension was confirmed.
Also, we reveal that, even with the assistance of ILRMA-based initialization, the SDRs of the conventional MNMFs and FastMNMFs with the full-rank spatial model cannot reach that of the proposed method.

As regards the optimization cost, the EM algorithm in the proposed method converged within five iterations, which was greatly reduced from the number of iterations required for MNMFs and FastMNMFs.
In addition, the actual computational times of MNMF, FastMNMF, and the proposed EM algorithm for each iteration were 10.18~s, 0.87~s, and 0.005~s, respectively, further illustrating the advantageousness of the proposed method.

On the other hand, the unbiased sample standard deviations of SDR improvements just after 200 iterations of ILRMA, original MNMF, ILRMA+MNMF, original FastMNMF, ILRMA+FastMNMF, and the proposed method are 0.19, 2.37, 0.38, 5.75, 0.26, and 0.22, respectively.
This means that the proposed method is a more stable algorithm than MNMF and FastMNMF in terms of initialization dependency.

\section{Conclusion}
We proposed a new algorithm that accurately and efficiently extracts a directional target source in diffuse background noise.
The proposed method is based on ILRMA and restores the lost spatial basis by using the EM algorithm to extend the spatial covariance of the noise from a rank-($M\!-\!1$) matrix to the full-rank matrix.
In an experiment, we confirmed that the proposed method outperforms the conventional methods in terms of accuracy and computational efficiency.

\section*{Acknowledgment}
This work was partly supported by SECOM Science and Technology Foundation and JSPS KAKENHI Grant Numbers 17H06101, 19H01116, and 19K20306.

\bibliographystyle{IEEEbib}
\bibliography{refs}

\end{document}